\documentclass[aps,twocolumn]{revtex4}
\usepackage{graphicx}
\usepackage{psfrag}
\usepackage{fancyhdr}

\begin{document}
\date{\today}

\title{Collisionless relaxation in non-neutral plasmas}

\author {Yan Levin, Renato Pakter, and Tarcisio N. Teles}

\address{
  Instituto de F\'{\i}sica, UFRGS, 
  Caixa Postal 15051, CEP 91501-970, Porto Alegre, RS, Brazil 
}  

\begin{abstract}
A theoretical framework is presented which allows to quantitatively predict
the final stationary state achieved by a non-neutral plasma  
during a process of collisionless relaxation. As a specific application, the theory
is  used to study relaxation of charged-particles beams.  
It is shown that a fully matched beam relaxes
to the Lynden-Bell distribution.  However, when a mismatch is present and the
beam  oscillates, parametric resonances
lead to a core-halo phase separation. The approach developed 
accounts for both the density and the velocity
distributions in the final stationary state.

\end{abstract}


\maketitle

Relaxation to a final stationary state (SS) of particles
interacting through long-range forces, such as (unscreened)
Coulomb or gravitational, is intrinsically different than that 
of systems with short-range interactions~\cite{Pa90}.  
In the latter
case, the interparticle collisions drive the system to an equilibrium
state described by the Maxwell-Boltzmann distribution. This
distribution is unique, in the sense that it is completely determined by the
globally conserved quantities such as the total 
energy, momentum, angular momentum, etc. ---  and is otherwise independent of specific
initial conditions. The same is true for neutral plasmas for which
the bare Coulomb potential is dynamically screened by the 
counter-charges, leading to a
well defined thermodynamic limit and equilibrium~\cite{Le02}.   
Relaxation of particles interacting by 
long-range (unscreened) potentials, on the other hand,
is very different.  For these systems, the 
collision duration time diverges and the state of
thermodynamic equilibrium is never reached.  
Instead, the dynamics evolves to a 
stationary state in which distribution functions 
{\it appear} to stop varying with time. Unlike thermodynamic equilibrium, in SS, however,
detailed balance is violated~\cite{Kl54} 
and neither {\it equilibrium} thermodynamics nor {\it equilibrium} statistical
mechanics can be used.   

In the limit in which the number of particles goes to infinity $(N \rightarrow \infty)$
while the total mass and 
charge are fixed, the non-neutral plasma is described {\it exactly} 
by the Vlasov equation~\cite{Br77},
\begin{equation}
 \label{e1}
 \frac{{\rm D}f}{{\rm D}t} \equiv {\partial f \over \partial t} + {\bf v} \cdot \nabla f 
+ {\bf F} \cdot \nabla_{\bf v} f =0, \\
\end{equation}
where ${\rm D}$ is the advective derivative,
$f(t,{\bf r},{\bf v})$ is the one particle distribution function, and 
${\bf F}$ is the mean force felt by particles at position ${\bf r}$. For simplicity
we have set particle mass to unity. 
Vlasov equation shows
that the distribution function evolves in time as an incompressible fluid.  If we now
discretize the height of the initial distribution function $f_0({\bf r},{\bf v})$ 
into a set of levels $\eta_j$, with $j=1...p$, 
the Vlasov dynamics of a $d$ dimensional system 
preserves each level's hypervolume 
$\gamma(\eta_j)=\int \delta(f(t,{\bf r},{\bf v})-\eta_j) {\rm d^d{\bf r}}{\rm d^d{\bf v}}$. 
For a general distribution function this
condition is equivalent to existence of an infinite number of dynamical 
invariants called the Casimir integrals or simply the Casimirs~\cite{Ch06}. 
One of the Casimirs  is the Boltzmann entropy which is, therefore, a constant of motion.
While the fine-grained
distribution function $f(t, {\bf r},{\bf v})$ never reaches a stationary state  -- the
evolution continues on smaller and smaller length scales {\it ad infinitum} --   
Lynden-Bell  argued that the {\it coarse-grained}
distribution function $\bar f(t, {\bf r},{\bf v},)$, {\it averaged} on microscopic length scales,
will rapidly relax 
to a meta-equilibrium with $\bar f({\bf r},{\bf v})$.  For gravitational systems Lynden-Bell called this process 
``a violent relaxation''~\cite{Ly67}.  
To obtain the stationary distribution $\bar f({\bf r},{\bf v})$, we divide the  
phase space into macrocells of volume ${\rm d^d{\bf r}}\,{\rm d^d{\bf v}}$, which
are in turn subdivided into $\nu$ microcells, each of volume $h^d$.  
As a consequence of incompressibility, each microcell can contain
at most one discretized level $\eta_j$.  The number density of the level $j$ 
inside a {\it macrocell}
at $({\bf r},{\bf v})$ --- 
number of microcells occupied by the level $j$ divided by $\nu$ --- 
will be denoted by $\rho_j({\bf r},{\bf v})$. 
Note that by construction,
the total number density of {\it all} levels in a macrocell 
is restricted to be 
\begin{equation}
 \label{e1a}
\sum_j \rho_j({\bf r}, {\bf v}) \le 1 \;.
\end{equation}
Using a standard combinatorial procedure~\cite{Ly67} it is then possible to associate
a coarse-grained entropy with the distribution of $\{\rho_j\}$. 
The entropy is found to be that of a $p$ species lattice gas,
\begin{eqnarray}
\label{e2}
S&=&-k_B \int \frac{{\rm d^d{\bf r}}\,{\rm d^d{\bf v}}}{h^d}  \left\{ \sum_{j=1}^p \rho_j ({\bf r},{\bf v}) \ln[\rho_j ({\bf r},{\bf v})] \right.\\ \nonumber
 &+& \left.[1-\sum_{j=1}^p\rho_j  ({\bf r},{\bf v})] \ln[1-\sum_{j=1}^p\rho_j ({\bf r},{\bf v})] \right\} \;,
\end{eqnarray}
where $k_B$ is the Boltzmann constant.   
Lynden-Bell argued that collisionless relaxation
should lead to the density distribution of levels which is the most likely, i.e. the
one that maximizes the {\it coarse-grained} entropy, consistent with the
conservation of all the dynamical 
invariants --- energy, momentum, angular momentum and the hypervolumes $\gamma(\eta_j)$.
In terms of the number densities $\{\rho_j\}$ which maximize the coarse-grained entropy, 
the stationary distribution function becomes
$\bar f({\bf r},{\bf v})=\sum_j \eta_j \rho_j({\bf r},{\bf v})$. 
The maximum entropy state, however, can only be 
achieved if there is a sufficient ergodicity (mixing)
in the phase space.

If the initial distribution $f_0({\bf r},{\bf v})$ 
is uniform ($p=1$), the maximization procedure
is particularly simple, yielding a Fermi-Dirac distribution,
\begin{eqnarray}
\label{e3}
\bar f({\bf r},{\bf v})=\eta_1 \rho({\bf r},{\bf v})=
\frac{\eta_1}{e^{\beta [\epsilon({\bf r},{\bf v})-\mu]}+1}\;,
\end{eqnarray}
where $\epsilon$ is the mean energy of particles with velocity ${\bf v}$
at position  ${\bf r}$, and 
$\beta$ and $\mu$ are the two Lagrange multipliers required by the conservations of
energy and number of particles,
\begin{eqnarray}
\label{e4}
\int{\rm d^d{\bf r}}\,{\rm d^d{\bf v}}\;\epsilon({\bf r},{\bf v}) \bar f({\bf r},{\bf v})=\epsilon_0 \\ \nonumber 
\int{\rm d^d{\bf r}}\,{\rm d^d{\bf v}} \bar f({\bf r},{\bf v})=1\;.
\end{eqnarray}
In the above formula 
$\epsilon_0$ is the energy per particle specified by the original distribution $f_0$.
For an azimuthally symmetric system, the mean  particle  energy  $\epsilon$ is 
a function of only the modulus $r$ and $v$. 
By analogy with the usual Fermi-Dirac statistics,
we define $\beta=1/k_B T$, were $T$ is the effective temperature of the stationary state
(not to be confused
with the usual definition of temperature in terms of the average kinetic energy which is 
valid only for classical systems in thermodynamic equilibrium) and 
$\mu$ is the effective plasma chemical potential.

In this Letter, we will show that when applied to non-oscillating confined non-neutral plasmas,
Eq.~(\ref{e3}) describes very accurately the final stationary state. 
On the other hand, if during the relaxation dynamics 
plasma undergoes collective oscillation, the theory of violent relaxation
fails dramatically.  Instead, we observe that
the system separates into two coexisting phases --- a cold core surrounded by a halo
of highly energetic particles.  The relaxation process is extremely
slow, taking tens of thousands of plasma oscillations to reach the stationary state. 
A new approach will then be presented which quantitatively predicts the phase-space 
distribution functions in the final relaxed state.

To illustrate the general theory, we will apply it to study the transport of
intense, continuous, charged-particles beams through a uniform focusing 
magnetic field~\cite{Dav01}. The beam is assumed to propagate with a constant axial
velocity $v_z{\bf \hat e}_z$, so that the axial coordinate $s=z=v_z t$
plays the role of time. 
The external focusing field is 
given by ${\bf B}=B_o{\bf \hat e}_z$ and is used to compensate
the repulsive Coulomb force  between the beam particles.
It is convenient to work in the Larmor frame of reference \cite{Dav01}, which
rotates with respect to the laboratory frame with angular
velocity $\Omega_L=qB_o/2\gamma_bmc$,
where $c$ is the speed of light in {\it vacuo}, 
and $q$, $m$, and $\gamma_b=[1-(v_z/c)^2]^{-1/2}$ are
the charge, mass, and relativistic factor of
the beam particles, respectively.
In this frame, the beam distribution 
function $f_b(s,{\bf r},{\bf v})$ evolves according 
to the Vlasov-Poisson system~\cite{Dav01}
\begin{eqnarray}
{\partial f_b \over \partial s} + {\bf v} \cdot \nabla f_b +(-\kappa_z {\bf r}
-\nabla
\psi) \cdot \nabla_{\bf v} f_b = 0, \label{e5} \\
\nabla^2 \psi = - (2 \pi K/N_b) \> n_b({\bf r},s), \label{e6}
\end{eqnarray}
where $N_b$ is the number of particles per unit axial length, 
${\bf r}$ is the position vector in the transverse plane, 
and ${\bf v}\equiv d{\bf r}/ds$ is the transverse velocity, 
$n_b({\bf r},s)=N_b\int f_b\, {\rm d}^2{\bf v}$ is the transverse beam density profile,
$\kappa_z=q^2 B^2_o/4\gamma_b^2 v_z^2 m^2 c^2$
is the focusing field parameter, and 
$K=2q^2N_b/\gamma_b^3 v_z^2m c^2$ is
the beam perveance, which is a measure of the beam intensity. 
In Eqs. (\ref{e5}) and (\ref{e6}),
$\psi$ is a scalar potential that incorporates both
self-electric and self-magnetic fields, ${\bf E}^s$ and
${\bf B}^s$.  We shall take zero
of the scalar potential to be at $r_w$, the position of the conducting channel wall.
The distribution function is 
normalized, so that $\int f_b\,{\rm d^2{\bf r}}\,{\rm d^2{\bf v}}=1$. 
In the Larmor frame, the system corresponds to 
a two dimensional non-neutral plasma
of pseudo particles of mass $m_p=1$ and charge $q=\sqrt{K/N_b}$
interacting by a repulsive logarithmic potential 
$\varphi(r)=-q^2 \ln(r/r_w)$, 
confined in a parabolic potential well of $U(r)=\kappa_z r^2/2$. We will
now explore the relaxation of these particles from the initially uniform distribution $(p=1)$, 
\begin{eqnarray}
\label{e7}
f_0({\bf r},{\bf v})=\eta_1 \Theta(r_m-r)\Theta(v_m-v)\;
\end{eqnarray}
with $\eta_1=1/\pi^2 r_m^2 v_m^2$, to the final stationary state.

At time $t=0$ the particles are uniformly distributed in the
phase space between $r\le r_m$ and $p \le p_m$.  
The distribution function Eq.~(\ref{e7}), 
however, is not a stationary solution of the Vlasov-Poisson system, and
for $t>0$
the system will start evolve in time.  
It is possible to adjust the values of $r_m$ and $v_m$
in such a way that during the evolution, the beam 
envelope (rms particle position) oscillates as little as possible. 
This corresponds to the so called matched beam 
condition --- the beam relaxes to equilibrium,
but without undergoing significant macroscopic oscillations.  
For the distribution function (\ref{e7}), the matching condition can be determined using
the beam envelope equation~\cite{Dav01,Dav97}. It is possible to show 
that the beam  will oscillate only little if 
$v_m^2 \approx \kappa_z r_m^2 -K$. 
When this  condition is met, we expect the mixing to be efficient  
and Lynden-Bell theory to apply.   The coarse-grained beam distribution 
should then relax to Eq. (\ref{e3}), 
with $\epsilon(r,v)=v^2/2 +U(r)+\psi(r)$, where the mean electrostatic potential $\psi(r)$ 
is determined self-consistently by an iterative solution of Eq.(\ref{e6}),
subject to constraints of Eqs. (\ref{e4}) with energy per particle given by
\begin{eqnarray}
\label{e8}
\epsilon_0=\frac{v_m^2}{4}+ \frac{\kappa_z r_m^2}{4}+\frac{1}{8}-\frac{K}{2}\ln\left(\frac{r_m}{r_w}\right)\;.
\end{eqnarray}
To compare the theory with the 
simulations, we calculate the number 
particles inside shells located between $r$ and $r+{\rm d}r$,
$N(r){\rm d}r=2 \pi N_b r {\rm d}r \int{\rm d^2{\bf v}} \bar f({\bf r},{\bf v})$; 
and the number of particles with velocities between $v$ and $v+{\rm d}v$,
$N(v){\rm d}v=2 \pi N_b v {\rm d}v \int{\rm d^2{\bf r}} \bar f({\bf r},{\bf v})$.
In Fig. 1 the solid lines show the values of $N(r)/N_b$ and $N(v)/N_b$ 
obtained using the theory
described above, while points are the result of a self-consistent N-particle 
dynamics simulation~\cite{NuPa07}. In all the figures distances are measured in 
units of $\sqrt{K/\kappa_z}$ and 
velocities in units of $\sqrt{K}$.
Excellent agreement between the theory and the simulation is found for
both position and velocity distributions 
{\it without} any fitting parameters.  We have checked that agreement
persists for other values of $r_m$ and $v_m$, as long as the matching condition is satisfied.   
The agreement, however, 
disappears as soon as the matching condition is violated and the beam begins 
to oscillate, Figs. (\ref{fig2},\ref{fig3}).
\begin{figure}[h]
\begin{center}
\includegraphics[width=8cm]{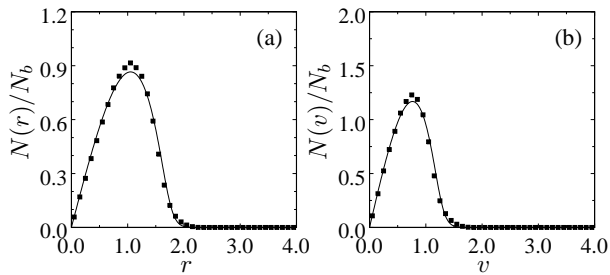}
\end{center}
\psfrag{r}{$r/\sqrt{K/\kappa_z}$}
\caption{Position and velocity distributions for a matched beam with 
$r_m=1.48 \sqrt{K/\kappa_z}$ and $v_m=1.1 \sqrt{K}$.  Solid line is the
theoretical prediction obtained using distribution function of Eq.~(\ref{e3}), 
while points are the result of dynamics simulation with
$N=5000$ particles.}
\label{fig1}
\end{figure}

Plasma oscillations lead to a number of important consequences
which are not taken into account in the theory of violent relaxation.  
For space-charge dominated inhomogeneous beams, the oscillations result 
in propagating density waves which
eventually break, emitting high energy particle jets~\cite{RiPa07}. 
The oscillations also excite  parametric resonances~\cite{Glu94} transferring
large amounts of energy to some particles at the expense of the
rest~\cite{NuPa07}, see Fig. 2.  Both of these mechanisms lead to 
inefficient phase space mixing and non-ergodicity.
\begin{figure}[h]
\begin{center}
\includegraphics[width=8cm]{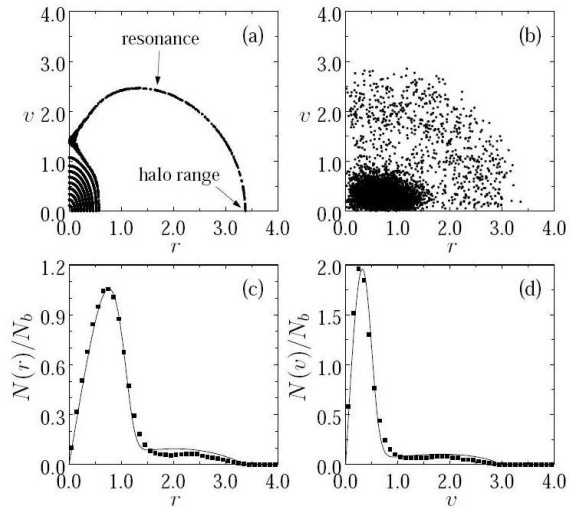}
\end{center}
\caption{  (a) Poincare section of a test particle moving in an oscillating potential controlled by
the envelope equation. The outermost curve shows the resonance which 
determines the halo's range. 
(b) A snapshot of the phase space 
particle distribution obtained using dynamics simulations  with $N=5000$,
after $40$ thousand beam
envelope oscillations. The halo particles are almost uniformly 
distributed in the phase space.  The halo's range $r_R$  is 
determined by the original parametric resonance, see  panel (a).
(c) Position and (d) velocity distributions.  Solid curves are the theoretical 
predictions obtained using the distribution function of 
Eq.~(\ref{e9}) and points are the results of dynamics simulations.
The initial distribution is uniform with  
$r_m=1.98 \sqrt{K/\kappa_z}$ and $v_m=0.24\sqrt{K}$.  It has exactly the same energy
as the fully matched distribution of Fig.\ref{fig1}, showing that SS depends
explicitly on the initial distribution.}
\label{fig2}
\end{figure}
As the relaxation proceeds, the oscillating beam core becomes
progressively colder, while a halo of highly energetic particles is created. 
Because of the incompressibility restriction imposed by the Vlasov
dynamics Eq.~(\ref{e1a}),
the core, however, can not freeze  --  collapse
to the minimum of the potential energy.  Instead, 
the distribution function of the core particles progressively 
approaches that of a fully degenerate Fermi gas.  

The extent of the halo is determined by the
location of the parametric resonance, and its range can be 
calculated using the canonical perturbation theory~\cite{Glu94}.  In Fig. 2a we show the Poincare
section of a test particle moving under the action of an oscillating beam potential calculated
using the envelope equation~\cite{NuPa07,RiEl95}.
The resonant orbit is the outermost curve of the 
Poincare plot.  The first resonant particles
move in almost a simple harmonic motion
with energy 
$\epsilon_R=-K \ln(r_R/r_w)+\kappa_z r_R^2/2$, where $r_R$ is the intersection of the
resonant trajectory with the $v=0$ axis.  As
more and more particles are ejected from the beam core their motion, however, 
becomes chaotic
and a halo distribution becomes smeared out.  We find that the distribution function
of a completely relaxed halo is very well approximated by the Heaviside step function
$\Theta(\epsilon_R-\epsilon)$.  

For an out of (thermodynamic) equilibrium system, there are no clear
parameters which will control the core-halo coexistence --- such as
pressure, temperature and chemical 
potentials for usual thermodynamic systems in coexistence.  
We can not, therefore,  {\it a priory} say
when the  halo formation will stop and a stationary state  
be established.  Empirically, however, we have observed that 
this happens when the core temperature becomes sufficiently low.  
In all cases studied, we find that the core-halo equilibrium is achieved when
the ratio between the core temperature and the corresponding Fermi temperature 
is $T/T_F \approx 1/40$ -- i.e. when
$\beta \mu \approx 40$. The distribution function for the 
core-halo system, then, takes a very simple form  
\begin{eqnarray}
\label{e9}
f_b({\bf r},{\bf v})=
\frac{\eta_1}{e^{\beta \epsilon({\bf r},{\bf v})-40}+1}+\chi \Theta(\epsilon_R-\epsilon)\;.
\end{eqnarray}
Since all the dependence on ${\bf r}$ and ${\bf v}$ enters only implicitly through
$\epsilon$, $f_b$ automatically satisfies the Vlasov-Poisson system.  The value of
$\eta_1=1/\pi^2 r_m^2 v_m^2$, is determined by the initial distribution $f_0$, while the value of 
$\epsilon_R$ is calculated from the location of the parametric resonance, Fig. 2a.  
This leaves to determine self-consistently, using Eqs.~(\ref{e4}) and (\ref{e6}), 
the mean electrostatic potential $\psi(r)$,
the inverse temperature  $\beta$, and the amplitude $\chi$ 
which will determine the fraction of particles inside the halo.  
These can, once again, be obtained 
iteratively.
\begin{figure}[h]
\begin{center}
\includegraphics[width=8cm]{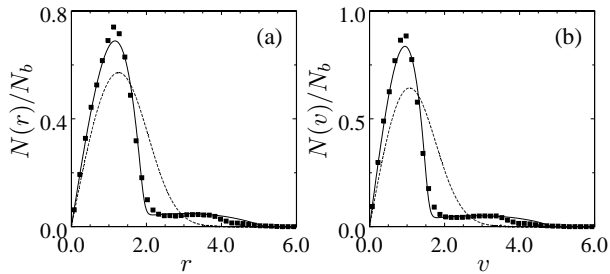}
\end{center}
\caption{(a) Position and (b) velocity distributions. 
Points are the result of  dynamics simulations.  
Solid curves are the theoretical 
predictions obtained using the distribution function of Eq.~(\ref{e9}). Dashed curves are
the predictions of the violent relaxation theory based on Eq.~(\ref{e3}).  The figure demonstrates
that for oscillating beams, mixing is inefficient and violent relaxation theory does not apply.  
The initial distribution is uniform with  
$r_m=1.0\sqrt{K/\kappa_z}$ and $v_m=2.4\sqrt{K}$.}
\label{fig3}
\end{figure}
In Figs. \ref{fig2} and \ref{fig3} 
we plot $N(r)/N_b$ and $N(v)/N_b$, obtained using the theory
presented above for two core-halo systems characterized by different values of initial
$r_m$ and $v_m$, and compare these distributions with the ones obtained using the  
dynamics simulations. Excellent agreement is found in all cases. 
In Fig. \ref{fig3} we
also present the distribution functions obtained 
using the violent relaxation theory, Eq. (\ref{e3}). 
It is clear that this theory is unable to describe relaxation of oscillating plasmas.

Up to now we have considered plasmas which at $t=0$ where uniformly distributed.
This, however, is not very usual in practice and more   
realistically one might expect a initially thermallized distribution of the form
\begin{equation}
\label{e10}
f_0({\bf r},{\bf v})=\frac{1}{2 \pi^2 \sigma^2 r_m^2 }\Theta(r_m-r) e^{-\frac{v^2}{2 \sigma^2}}\;.
\end{equation}
The procedure is then to discretize the Gaussian part of 
this distribution into $p$ levels.  At the lowest order, we
can take $p=1$ and approximate Eq.~(\ref{e10}) by Eq.~(\ref{e7}).  
To have equal energy, 
both distributions must have the same values of $\langle v^2 \rangle$.
This requires that $v_m=2 \sigma$. The final relaxed distribution of this core-halo
system should then be given by Eq.~(\ref{e9}) with $\eta_1=1/4 \pi^2 r_m^2 \sigma^2$.
In Fig. 4 we plot $N(r)/N_b$ and $N(v)/N_b$  and 
compare these distributions with the ones obtained using the
dynamics simulation in which the initial particle  
positions and velocities were distributed according to
Eq.(\ref{e10}).  In spite of the crudeness of the
one level approximation, an amazingly good agreement between the theory 
and the simulations
is obtained {\it without any adjustable parameters}.
We have checked that this good agreement persists for other values
of $\sigma$ and $r_m$. 

\begin{figure}[h] 
\begin{center}
\includegraphics[width=8cm]{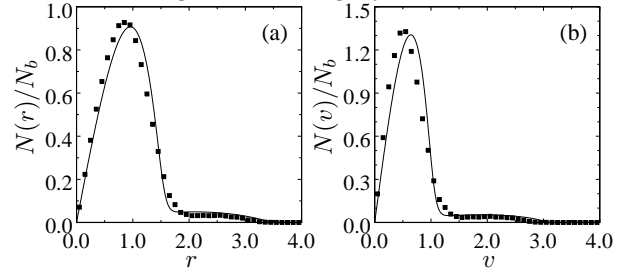}
\end{center}
\caption{(a) Position and (b) velocity  distributions.  
Solid curves are the theoretical 
predictions and points are the result of dynamics simulations. 
Initial $t=0$ distribution is a thermal one given by Eq.~(\ref{e10})  
$r_m=1.0\sqrt{K/\kappa_z} $ and $\sigma=0.64\sqrt{K}$.}
\label{fig4}
\end{figure}

In this Letter we have studied confined one component plasmas of charges  
interacting by unscreened 
Coulomb potential.  Unlike normal gases with short range forces,
non-neutral plasmas do not evolve to the state of thermodynamic equilibrium.  
Instead collisionless relaxation
culminates in a stationary state in 
which the detailed balance is violated.   Using a combination
of non-equilibrium statistical mechanics and the theory of parametric resonances
it is, nevertheless, possible to {\it a priory} predict the distribution functions 
for the final stationary state. 
Unlike the normal  thermodynamic
equilibrium, this state, however, explicitly depends  
on the initial distribution of particle velocities and positions.  

We would like to thank Felipe Rizzato for interesting discussions. 
This work is partially supported by CNPq and by the 
US-AFOSR under the grant FA9550-06-1-0345.


\end{document}